\documentclass[article]{IEEEtran}
\usepackage{cite}
\usepackage{booktabs}
\usepackage{multirow}
\usepackage{amsmath}
\usepackage{array}
\usepackage{amssymb}
\usepackage{bm}
\usepackage{mathrsfs}
\usepackage{ifpdf}
\usepackage{cite}
\usepackage{graphicx}
\usepackage{amsmath}
\usepackage{algorithmic}
\usepackage{dblfloatfix}
\usepackage{array}
\usepackage{fixltx2e}
\usepackage{mathrsfs,multirow,makecell}
\usepackage{url}
\usepackage{longtable}
\usepackage[comma,numbers,square,sort&compress]{natbib}
\usepackage{cases}
\usepackage{microtype}
\usepackage{color}
\usepackage{epstopdf}
\usepackage{float}
\usepackage{subfigure}
\usepackage{graphics} % for pdf, bitmapped graphics files
\usepackage{epsfig} % for postscript graphics files
\usepackage{mathptmx} % assumes new font selection scheme installed
\usepackage{times} % assumes new font selection scheme installed
\usepackage{amssymb}  % assumes amsmath package installed
\usepackage{amstext}
\usepackage{amsfonts}
\usepackage{amsbsy}
\begin{document}

\newtheorem{Theorem}{Theorem}
\newtheorem{Assumption}{Assumption}
\newtheorem{Proposition}{Proposition}
\newtheorem{Definition}{Definition}
\newtheorem{Lemma}{Lemma}
\newtheorem{Condition}{Condition}
\newtheorem{Corollary}{Corollary}
\newtheorem{Conjecture}{Conjecture}
\newtheorem{Remark}{Remark}
\newtheorem{Example}{Example}
\newtheorem{Step}{Step}
\newtheorem{Case}{Case}
\newtheorem{Proof}{Proof}

\title{The necessary and sufficient condition for the uncertain control gain in active disturbance rejection control}
\author{Sen Chen, Yi Huang$^*$, Zhiliang Zhao% <-this % stops a space
\thanks{Sen Chen and Zhiliang Zhao are with the School of Mathematics and Information Science, Shaanxi Normal University, Xi'an 710119, Shaanxi, China.
Yi Huang are with the Key Laboratory of Systems and Control, Academy of Mathematics and Systems Science, Chinese Academy of Sciences, Beijing 100190, P. R. China. (yhuang@amss.ac.cn)}}
\markboth{}%
{Shell \MakeLowercase{\textit{et al.}}: Bare Demo of IEEEtran.cls for IEEE Journals}
\maketitle
\begin{abstract}
Considering the control problem for nonlinear uncertain systems, the tolerable range of uncertain control input gain is a fundamental issue. The paper presents the necessary and sufficient condition for the well-performed closed-loop system based on active disturbance rejection control (ADRC) design. Via the proposed necessary and sufficient condition, the maximum tolerable range of the uncertain control gain is quantitatively presented, which reveals the capability of ADRC to handle the uncertainties related with control input. Moreover, under the proposed necessary and sufficient condition, both the transient performance and the steady state property of the ADRC based closed-loop system are rigorously analyzed.
\end{abstract}

\IEEEpeerreviewmaketitle

\section{Introduction}

How to deal with uncertainties and/or disturbances is a central issue in control technology and control science \cite{kumar2014control}.
Plenty of methods with the aim to ensure normal operation of systems against uncertainties and/or disturbances have been substantially developed, such as proportional-integral-derivative (PID) control \cite{Wang2007PIDTAC}, adaptive control \cite{Li2014RobustAdapTAC}, sliding-mode control \cite{Zhang2015slidingTAC}, and various disturbance rejection methods including disturbance observer based control (DOBC) \cite{Chen2015DOTAC, Chen7265050TIE}, extended high gain observer based control (EHGOBC) \cite{freidovich2008performance, khalil2017cascadeAc, khalil2017highBOOK}, and active disturbance rejection control (ADRC)  \cite{han2009pid, HuangADRCreviewISA, Guo2015ADRCTAC}.

The disturbance rejection based control approaches are featured with two degrees of freedom (2-DOF) design, i.e., one to achieve estimation and compensation for uncertainties and/or disturbances, and the other to force the closed-loop system to attain the desired performance.
In the last decades, the disturbance rejection based control approaches have been widely implemented in industrial applications due to the simplicity in engineering implementation and superior transient performance \cite{Ohnishi6822593TIE, Texas2013, sun2016tuning, Chen7265050TIE}.

Inspired by the successful applications, the theoretical studies of the disturbance rejection methods have attracted increasing attentions.
Numerous researches reveal the strong capability of the disturbance rejection methods to handle a large scope of the uncertainties, which are related with the system states and the time, no matter linear or nonlinear, continuous or discontinuous, matched or mismatched \cite{HuangADRCreviewISA, xue2018performance, Zhao7959174Tac, guo2015active}.
However, the uncertainties related with the control input lead to the unknown coupled nonlinearities affecting both the system dynamics and the estimating process, which might destroy the stability of the closed-loop systems.
Moreover, when there exist uncertainties related with control input, the stability analysis of disturbance rejection methods becomes extremely difficult.
The studies of the maximum range of the uncertain control gain are progressing haltingly.
Many studies neglect the uncertainties related with the control input \cite{WANG2016AC, Zhao7959174Tac}, or just assume that the lumped disturbance satisfies some bounded conditions \cite{Yao7900325TIE}.
Some researches rigorously study the stability condition of the uncertain control gain \cite{freidovich2008performance, Ohnishi6822593TIE, xue2018performance, guo2015active}.
The small gain theorem is utilized to analyze the stability of the DOBC based closed-loop system, where the detailed condition of the uncertain control gain is masked and hard to be check in practice \cite{Ohnishi6822593TIE}.
With the precondition of the existence of some Lyapunov functions, the tolerable range of the uncertain control gain for the nonlinear ADRC is related to the Lyapunov functions \cite{guo2015active}, which is difficult to verify in practice.
By small gain theorem, the boundary of the uncertain control gain for the EHGOBC is expressed as a function of the observer parameters \cite{freidovich2008performance}.
Based on circle criterion, \cite{xue2018performance} presents an explicit range of uncertain control gain for the linear ADRC.
Nevertheless, the existing studies \cite{freidovich2008performance, Ohnishi6822593TIE, xue2018performance, guo2015active} only present the sufficient stability conditions of the uncertain control gain.
Thus, the following question rises \cite{Chen7265050TIE}.

\emph{What is the maximum range of the uncertain control gain for the disturbance rejection methods?}

Motivated by this fundamental problem of the disturbance rejection methods, the paper studies the capability to handle the uncertain control gain for a typical disturbance rejection design, i.e., ADRC. Via the estimation of the total disturbance from the extended state observer (ESO), the ADRC design, which contains the compensation for the total disturbance and the control law for the system in the ideal integrator chain form, is proposed in \cite{han2009pid}.
The paper analyzes the necessary and sufficient condition under which the ADRC based closed-loop system has satisfied tracking performance that the tracking error can be tuned in a small region by the ESO's parameters. Furthermore, the proposed necessary and sufficient condition demonstrates the maximum range of the uncertain control gain, which can be further expressed in the explicit form for systems with the dimension less than five.
Moreover, under the proposed necessary and sufficient condition, both the transient performance and the steady state property of the ADRC are rigorously analyzed.
%Additionally, the simulation results show the feasibility of the theoretical analysis in the paper. Finally, a tuning law to extend the maximum range of the uncertain control gain is further tested.
The main contributions of the paper are shown as follows.

(i). The necessary and sufficient condition for the well-performed closed-loop based on ADRC is presented, which further reveals the maximum capability to handle the uncertain control gain.

(ii). Via the proposed necessary and sufficient condition, the maximum range of the uncertain control gain is explicitly shown for systems with the dimension less than five.

(iii). Under the proposed necessary and sufficient condition, both the transient performance and the steady state property of the ADRC based closed-loop system are analyzed.

%(iv). The proposed necessary and sufficient condition further illuminates a tuning law to extend the maximum range of the uncertain control gain.

The rest of the paper has the following organization.
In Section 2, the problem formulation and the ADRC design are presented.
The main theoretical results are shown in Section 3.
In Section 4, the simulation studies are given.
The conclusion is presented in Section 5.

\section{Problem Formulation}

Consider the following class of nonlinear uncertain systems:
\begin{equation}\label{plant}
\left\{\begin{split}
&\dot{X}(t) = A X(t)+ B (\bar{b} u(t) + f(X,u,t)),\\
&y(t) = C^T X(t),
\end{split} \right.
\quad t\geq t_0.
\end{equation}
where $X(t) = [x_1(t)~x_2(t)~\cdots~x_n(t)]^T\in R^n$ is the system state, $u(t)\in R$ is the control input, $y(t)\in R$ is the measured output, and the matrices $(A,B,C)$ satisfy the following integrator chain form:
\begin{equation}\label{mtr_integrator}
A = \begin{bmatrix}0&1&\cdots& 0 \\ \vdots & \ddots & \ddots &  \vdots \\   0 & \cdots & 0 &   1 \\ 0 & \cdots  &\cdots & 0 \end{bmatrix}_{n \times n},
B = \begin{bmatrix} 0 \\ \vdots \\ 0 \\   1 \end{bmatrix}_{n \times 1},
C = \begin{bmatrix} 1 \\ 0 \\ \vdots \\ 0  \end{bmatrix}_{n \times 1}.
\end{equation}
The nonzero constant $\bar{b}$ is the known nominal control gain, and $f(X,u,t)$ is the total disturbance, which has the following form:
\begin{equation}
f(X,u,t) = g(X,t) +  b_\delta u(t),
\end{equation}
where the constant $ b_\delta $ is the uncertain control gain, and $g(X,t)$ represents the combination of the external disturbances, unmodeled dynamics, and  parametric perturbations, which satisfies the following assumption.

\begin{Assumption}\label{ass_uncertainty}
The function $g(X,t)$ is continuously differentiable, and satisfies the following equation.
\begin{equation}\label{AssUncertaintyBound}
\sup_{  t \in R}\left\{|g(X,t)|, \left\|\frac{\partial g(X,t)}{\partial X} \right\| , \left|\frac{\partial g(X,t)}{\partial t} \right|  \right\} \leq \psi(X),
\end{equation}
where $\psi$ is a continuous function.
\end{Assumption}
\begin{Remark}
Assumption~\ref{ass_uncertainty} describes the size of the total disturbance in (\ref{plant}), which contains a large scope of nonlinearities, including trigonometric functions, polynomial functions and exponential functions. Assumption~\ref{ass_uncertainty} is more general than the assumptions in \cite{xue2018performance, WANG2016AC}.
\end{Remark}

\begin{Remark}
For the sake of simplicity and clarity, the paper studies the integrator chain model  (\ref{plant})--(\ref{mtr_integrator}), which is the kernel of most uncertain nonlinear systems \cite{Zhao7959174Tac, Chen2019SCIS}.
\end{Remark}

The control objective for the uncertain system (\ref{plant}) is to design the control input $u(t)$ such that the state $X(t)$ can track the reference signal $R(t)= [r(t)~\dot{r}(t)~\cdots ~ r^{(n-1)}(t)]^T\in R^n$ under a class of uncertainties $f(X,u,t)$.
The reference signal $R(t)$ and its derivative are bounded, i.e., $\sup_{t\geq t_0,0\leq i\leq n+1} \{|r^{(i)}(t)|\} \leq \eta_r$ for some positive constant $\eta_r$.

To achieve the control objective, a commonly used ADRC approach, which is featured with the estimation and compensation for the total disturbance, is presented as follows.

The following ESO is presented to online estimate the system state $X$ and the total disturbance $f(X,u,t)$.
\begin{equation}\label{RESO}
\begin{bmatrix}\dot{\hat{X}}(t) \\ \dot{\hat{f}} (t) \end{bmatrix}
= A_e \begin{bmatrix}\hat{X}(t) \\ \hat{f} (t) \end{bmatrix} + B_e \bar{b} u(t) - L_e \left(C_e^T \begin{bmatrix}\hat{X}(t) \\ \hat{f} (t) \end{bmatrix} - y(t)\right),
\end{equation}
where $\hat{X}(t)=[\hat{x}_1(t) ~\cdots ~\hat{x}_n(t)]\in R^n$ is the estimation of the state $X(t)$, $\hat{f}(t)\in R$ is the estimation of the total disturbance $f(X,u,t)$, the matrices $(A_e,B_e,C_e)$ have the following form:
\begin{equation}\nonumber
A_e = \begin{bmatrix}A & B \\ 0 & 0 \end{bmatrix}_{(n+1)\times (n+1)}, B_e = \begin{bmatrix} B \\ 0 \end{bmatrix}_{(n+1)\times 1},
C_e = \begin{bmatrix} C \\ 0 \end{bmatrix}_{(n+1)\times 1},
\end{equation}
and $L_e=[l_1~ l_2~ \cdots ~l_{n+1}]^T\in R^{n+1}$ is the ESO's parameter to be designed such that $A_{Le} \triangleq A_e-L_eC_e$ is Hurwitz.

A popular way to design the ESO's parameter $L_e$ is shown as follows \cite{freidovich2008performance, khalil2017cascadeAc, khalil2017highBOOK, xue2018performance, GaoESObandwidth}:
\begin{equation}
l_i = \phi_i \omega_o^{i},\quad
1\leq i \leq n+1,
\end{equation}
where the parameters $\phi_i$ ensure that the polynomial $P(s)=s^{n+1}+\Sigma_{i=0}^{i=n}\phi_{n+1-i} s^{i}$ is Hurwitz.

Based on the online estimations $\hat{X}$ and $\hat{f}$ from the ESO (\ref{RESO}), the ADRC input composed of the disturbance rejection can be designed as follows:
\begin{equation}\label{control_ADRC}
u(t) = \left\{\begin{split}
&0 , \quad t_0 \leq t < t_u\\
&-\frac{\hat{f}(t)}{\bar{b}} +\frac{-K^T (\hat{X}(t)-R(t))+  r^{(n)}(t) }{\bar{b}}, \quad t\geq t_u,
\end{split}\right.
\end{equation}
where $K = [k_1~k_2~\cdots ~ k_n]^T \in R^n$ is the feedback gain to be designed such that $A_K \triangleq A-BK^T$ is Hurwitz, and $t_u$ is the time after which the peaking of the ESO (\ref{RESO}) ends, which can be designed as follows \cite{xue2018performance}:
\begin{equation}\label{tu}
t_u = t_0 +2(n-1) \|P_1\| \max \left\{ \frac{\ln(\omega_o \tilde{\rho}_0)}{\omega_o},~0 \right\},
\end{equation}
where the positive $\tilde{\rho}_0$ satisfies $\max_{2\leq i\leq n}|x_i(t_0)-\hat{x}_i(t_0)|\leq \tilde{\rho}_0$ and the positive-definite matrix $P_1$ satisfies
\begin{equation}\label{a1p1}
A_1^T P_1 + P_1 A_1 =-I, \quad
A_1 = \left[\begin{smallmatrix}-\phi_1&1&\cdots& 0 \\ -\phi_2 & 0 & \ddots &  \vdots \\   \vdots & \vdots & \ddots &   1 \\ -\phi_{n+1} & 0  &\cdots & 0 \end{smallmatrix}\right]_{(n+1) \times (n+1)}.
\end{equation}
It is remarkable that $t_u=t_0$ if the initial condition satisfies $\mathop{\max}\limits_{2\leq i\leq n}|x_i(t_0)-\hat{x}_i(t_0)|\leq \frac{1}{\omega_o} $.

\begin{Remark}\label{Remark_bandwidth}
By designing
\begin{equation}\label{bandwidthPHI}
\phi_i = \frac{(n+1-i)!i!}{(n+1)!},\quad
1\leq i \leq n+1,
\end{equation}
the polynomial $P(s)=s^{n+1}+\Sigma_{i=0}^{i=n}\phi_{n+1-i} s^{i}=(s+1)^{n+1}$, which implies that all the eigenvalues of $A_{Le}$ are $-\omega_o$, where $\omega_o>0$ is named as the bandwidth of the ESO \cite{GaoESObandwidth}.
\end{Remark}

\begin{Remark}\label{Remark_ADRCdesign}
The presented ADRC controller is a typical disturbance rejection design, which is also known as EHGOBC with linear extended high observer \cite{freidovich2008performance, khalil2017cascadeAc, khalil2017highBOOK}.
However, only the sufficient condition for uncertain control input gain is studied for this approach \cite{freidovich2008performance, xue2018performance, guo2015active}.
The paper aims to analyze the maximum range of uncertain control input gain for the ADRC design.
\end{Remark}

In the next section, the theoretical analysis of the ADRC controller (\ref{RESO})--(\ref{a1p1}), including the well-performed condition, transient performance and the steady state property, will be presented.

\section{Theoretical results}

\subsection{Well-performed condition and transient performance}

As shown in the existing study \cite{xue2018performance}, the transient performance of the ADRC based closed-loop system is satisfied if the uncertain control input gain satisfies some conditions. This subsection further investigates the necessary and sufficient condition of the uncertain control input gain for the satisfied transient performance.

Consider the ideal trajectory $X^*$, which satisfies the following dynamics:
\begin{equation}\label{idealTra}
\begin{split}
&\dot{X}^*(t) = AX^*(t)-BK^T(X^*(t)-R(t))+B r^{(n)}(t),\\
& X^*(t_0) = X(t_0).
%\\
%&y^*(t) = C^T X^*(t), \quad X^*(t_0) = X(t_0),
\end{split}
\end{equation}

\begin{Remark}
For the ideal system (\ref{idealTra}), the tracking error $|X^*(t)-R(t)|$ will globally exponentially converge to zero.
\end{Remark}

The following definition describes the property of the ADRC to be well-performed.

\begin{Definition}\label{Def1}
If for any given positive $\varepsilon$, there exists ADRC controller (\ref{RESO})--(\ref{a1p1}) such that the trajectory of the uncertain system (\ref{plant}) satisfies
\begin{equation}\label{ThmStableX*}
\sup_{t\in[t_0,\infty)} \| X(t)-X^*(t) \| \leq \varepsilon,
\end{equation}
then the ADRC (\ref{RESO})--(\ref{a1p1}) is defined to be well-performed.
\end{Definition}

In Definition~\ref{Def1}, (\ref{ThmStableX*}) means that the error between the actual trajectory and the ideal trajectory is sufficiently small, which meets the requirement in practice.
Then, the following theorem presents the necessary and sufficient condition of the uncertain control gain for the  ADRC (\ref{RESO})--(\ref{a1p1}) to be well-performed.

\begin{Theorem}\label{ThmStable}
Consider the system (\ref{plant}) and the ADRC controller (\ref{RESO})--(\ref{a1p1}) with Assumption~\ref{ass_uncertainty}. The following two statements are equivalent:

(S1). The matrix
\begin{equation}
A_2 = \left[\begin{smallmatrix}-\phi_1&1&\cdots& 0 \\ -\phi_2 & 0 & \ddots &  \vdots \\   \vdots & \vdots & \ddots &   1 \\ -\phi_{n+1} \left(1+ \frac{b_\delta}{\bar{b}}\right) & 0  &\cdots & 0 \end{smallmatrix}\right]_{(n+1) \times (n+1)}
\end{equation}
is Hurwitz.

(S2). For any given positive $\varepsilon$, any given initial values $(X(t_0),\hat{X}(t_0),\hat{f}(t_0),R(t_0))$ and $(\psi,K,\bar{b},b_\delta,\eta_r)$, there exists a positive $\omega_o$, such that the closed-loop system satisfies (\ref{ThmStableX*}).
\end{Theorem}

The proof of Theorem~\ref{ThmStable} is given in Appendix.

Theorem~\ref{ThmStable} demonstrates the necessary and sufficient condition for the well-performed closed-loop system (\ref{plant}) and (\ref{RESO})--(\ref{a1p1}), i.e., the matrix $A_2$ is Hurwitz. The stability of the matrix $A_2$ further reveals the capability of the ADRC to handle the uncertain control gain.

\begin{Remark}
For any given $\phi_i~(1\leq i \leq n)$, the maximum range of the uncertain control gain can be  calculated by Routh criterion.
\end{Remark}

Then, the comparison with the existing theoretical analysis on the uncertain control gain will be quantitatively presented.

The existing study of the ADRC (\ref{RESO})--(\ref{a1p1}) with the $\phi_i$ (\ref{bandwidthPHI}) is shown in the following lemma.

\begin{Lemma}\label{lem_conditionb}
\cite{xue2018performance}
Consider the system (\ref{plant}) and the ADRC controller (\ref{RESO})--(\ref{a1p1}) with Assumption~\ref{ass_uncertainty}. Let $\phi_i (1\leq i \leq n+1)$ satisfy (\ref{bandwidthPHI}) and $(b,b_\delta)$ satisfy $\frac{b_\delta}{\bar{b}} \in \left(-1,1+\frac{2}{n}\right)$.
Then, the closed-loop system satisfies
\begin{equation}
\sup_{t\in[t_0,\infty)} \|X(t)-X^*(t)\| \leq \tilde{\eta}_1^* \frac{\max \{\ln\omega_o, 1\}}{\omega_o},
\end{equation}
for all $\omega_o\geq \tilde{\omega}_o^*$, where $\tilde{\omega}_o^*$ and $\tilde{\eta}_1^*$ are positives dependent on $(K,\psi,\bar{b},b_\delta,r)$ and the bounds of initial values $(X(t_0),\hat{X}(t_0),\hat{f}(t_0),R(t_0))$.
\end{Lemma}

%\begin{table}[!t]
%\centering
%\footnotesize
%\caption{Ranges of the uncertainty of control gain.}
%\label{tab2}
%\tabcolsep 10pt %space between two columns.
%\begin{tabular*}{\textwidth}{ccccccccc}
%    \hline
%    $n$: & $1$ & $2$ & $3$ & $4$ & $5$ \\
%    \hline
%    $b_\delta/\bar{b}$ (Theorem~\ref{ThmStable}): & $(-1,\infty)$ & $(-1,8)$ & $(-1,4)$ & $(-1,4)$ & $(-1,2.3704)$ \\
%    \hline
%    $b_\delta/\bar{b}$ (Lemma~\ref{lem_conditionb}): & $(-1,3)$ & $(-1,2)$ & $(-1,\frac{5}{3})$ & $(-1,\frac{3}{2})$ & $(-1,\frac{7}{5})$ \\
%    \hline
%\end{tabular*}
%\end{table}

\begin{table}[!t]
	\centering
	\footnotesize
	\caption{Ranges of the uncertainty of control gain.}
	\label{tab2}
	%\centering
	%\tabcolsep 1pt %space between two columns. 用于调整列间距
	\begin{tabular}{ccc c c c}
		\hline
    $n$: & $1$ & $2$ & $3$ & $4$ & $5$ \\
    \hline
    $b_\delta/\bar{b}$ (Theorem~\ref{ThmStable}): & $(-1,\infty)$ & $(-1,8)$ & $(-1,4)$ & $(-1,4)$ & $(-1,2.37)$ \\
    \hline
    $b_\delta/\bar{b}$ (Lemma~\ref{lem_conditionb}): & $(-1,3)$ & $(-1,2)$ & $(-1,\frac{5}{3})$ & $(-1,\frac{3}{2})$ & $(-1,\frac{7}{5})$ \\
    \hline
	\end{tabular}
\end{table}

For the system order $n\in[1,5]$, the ranges of the uncertain control gain obtained from Theorem~\ref{ThmStable} and Lemma~\ref{lem_conditionb} are shown in Table~\ref{tab2}.
From Table~\ref{tab2}, the maximum range of the uncertain control gain obtained by the necessary and sufficient condition in Theorem~\ref{ThmStable} greatly improves the existing results.

\begin{Remark}\label{Prop_transient}
Under the conditions that the matrix $A_2$ is Hurwitz and Assumption~\ref{ass_uncertainty}, the following transient performance of the ADRC based closed-loop system (\ref{plant}) and (\ref{RESO})--(\ref{a1p1}) is shown:
\begin{align}
& \sup_{t\in [t_0,\infty)} \|X(t)-X^*(t)\| \leq \eta_1 \frac{\max\{\ln \omega_o , 1\}}{\omega_o},\label{Thm_re1_track}\\
& \left\| \begin{bmatrix} X(t)-\hat{X}(t) \\ f(X,u,t) - \hat{f}(t)  \end{bmatrix} \right\| \leq \eta_2\left( e^{\frac{- \omega_o (t-t_u)}{2}} +\frac{1}{\omega_o} \right),~~ t\geq t_u , \label{Thm_re2_est}
\end{align}
for $\omega_o\geq \omega_o^*$, where $(\omega_o^*,\eta_1,\eta_2)$ are positives dependent on $(\psi,K,\bar{b},b_\delta,\eta_r)$ and the bounds of initial values $(X(t_0),\hat{X}(t_0),\hat{f}(t_0),R(t_0))$.
The proof of (\ref{Thm_re1_track}) and (\ref{Thm_re2_est}) can be obtained via the similar proof of Theorem~\ref{ThmStable}. Therefore, both the tracking and the estimating error can be tunable by the ESO's parameter $\omega_o$.
\end{Remark}

\subsection{Steady state property}

In this subsection, the steady state property of the closed-loop system based on the proposed ADRC (\ref{RESO})--(\ref{a1p1}) is studied.

Assume that the reference signal and the uncertainties in the system (\ref{plant}) satisfy the following assumption.

\begin{Assumption}\label{ass_limit}
The total disturbance and reference signal satisfy
\begin{equation}\label{ass4}
\begin{split}
&g(X,t) = g_1(X)+d_1(t),~ ~\lim_{t\rightarrow \infty} d_1(t) \text{~~exists},\\
&\lim\limits_{t\rightarrow \infty} r^{i}(t)=0,~~1\leq i \leq n+1.
\end{split}
\end{equation}
\end{Assumption}

\begin{Remark}
Assumption~\ref{ass_limit} demonstrates that the external disturbances $d_1$ and the reference signal for the output $r$ converge to some constants, while there is no additional condition for the internal uncertainties $g_1$. Assumption~\ref{ass_limit} generalizes the existing conditions in \cite{WANG2016AC}, which takes no account of internal uncertainties.
\end{Remark}

The following theorem illustrates the steady state property of the ADRC (\ref{RESO})--(\ref{a1p1}) based closed-loop system.

\begin{Theorem}\label{Thm_converge}
Consider the ADRC based closed-loop system (\ref{plant}) and (\ref{RESO})--(\ref{a1p1}).
Let Assumption~\ref{ass_uncertainty}--\ref{ass_limit} be satisfied and the matrix $A_2$ be Hurwitz.
Then, there exists a positive $\omega_o^{**}$ dependent on the bounds of the initial values $(X(t_0),R(t_0),\hat{X}(t_0),\hat{f}(t_0))$ and $(\psi,K,b_\delta,\bar{b},\eta_r)$ such that for any $\omega_o \geq \omega_o^{**}$,
\begin{equation}\label{Thm_converge_re}
\left\{ \begin{split}
&\lim_{t\rightarrow \infty} (X(t)-R(t))=0,\\
&\lim_{t\rightarrow \infty} \begin{bmatrix}X(t) - \hat{X}(t) \\ f(X(t),u(t),t)-\hat{f}(t)\end{bmatrix}=0.
\end{split} \right.
\end{equation}
\end{Theorem}

The proof of Theorem~\ref{Thm_converge} is presented in Appendix.

Theorem~\ref{Thm_converge} demonstrates that the ADRC (\ref{RESO})--(\ref{a1p1}) based closed-loop system is semi-global asymptotically stable, if the external disturbance $d_1$ and the reference signal $r$ asymptotically approach to some constants.

\section{Simulation}

In this section, the simulations for the ADRC to deal with a large scope of the uncertain control gain are presented. Moreover, to extend the maximum range of the uncertain control gain, a tuning law of $\phi_i$ is proposed and tested.

Consider the uncertain system (\ref{plant}) with the system order $n=2$. The known nominal control gain is selected as  $\bar{b}=1$.
The reference signal is $R=\emph{\textbf{0}}$.

The following four groups of uncertainties, including parametric perturbations, nonlinear unmodeled dynamics, and external disturbances, are considered.
\begin{equation}\label{DifferentUncertainty}
\begin{split}
&\text{Case 1: } g(X,t) = 3x_1+3x_2;\quad\\
&\text{Case 2: } g(X,t) = 3x_1+x_1^2+3x_2+x_2^2;~\\
&\text{Case 3: } g(X,t) = 0.4\sin(x_1)+10\sin\left(\frac{\pi t}{8}\right) ;\\
&\text{Case 4: } g(X,t) =\left\{\begin{split}
& 0.1x_1,~t\in[0,5),\\
& 0.1x_1 + \frac{10(t-5)}{3},~t\in[5,8),\\
& 0.1x_1+ 10 ,~t\in[8,\infty).
\end{split}\right.
\end{split}
\end{equation}
The control parameters are chosen as $(k_1 ,k_2 , \omega_o)=(4,4,10000)$.

\begin{figure}[t]\centering
      \includegraphics[width=1\linewidth,trim=50 250 50 250,clip]{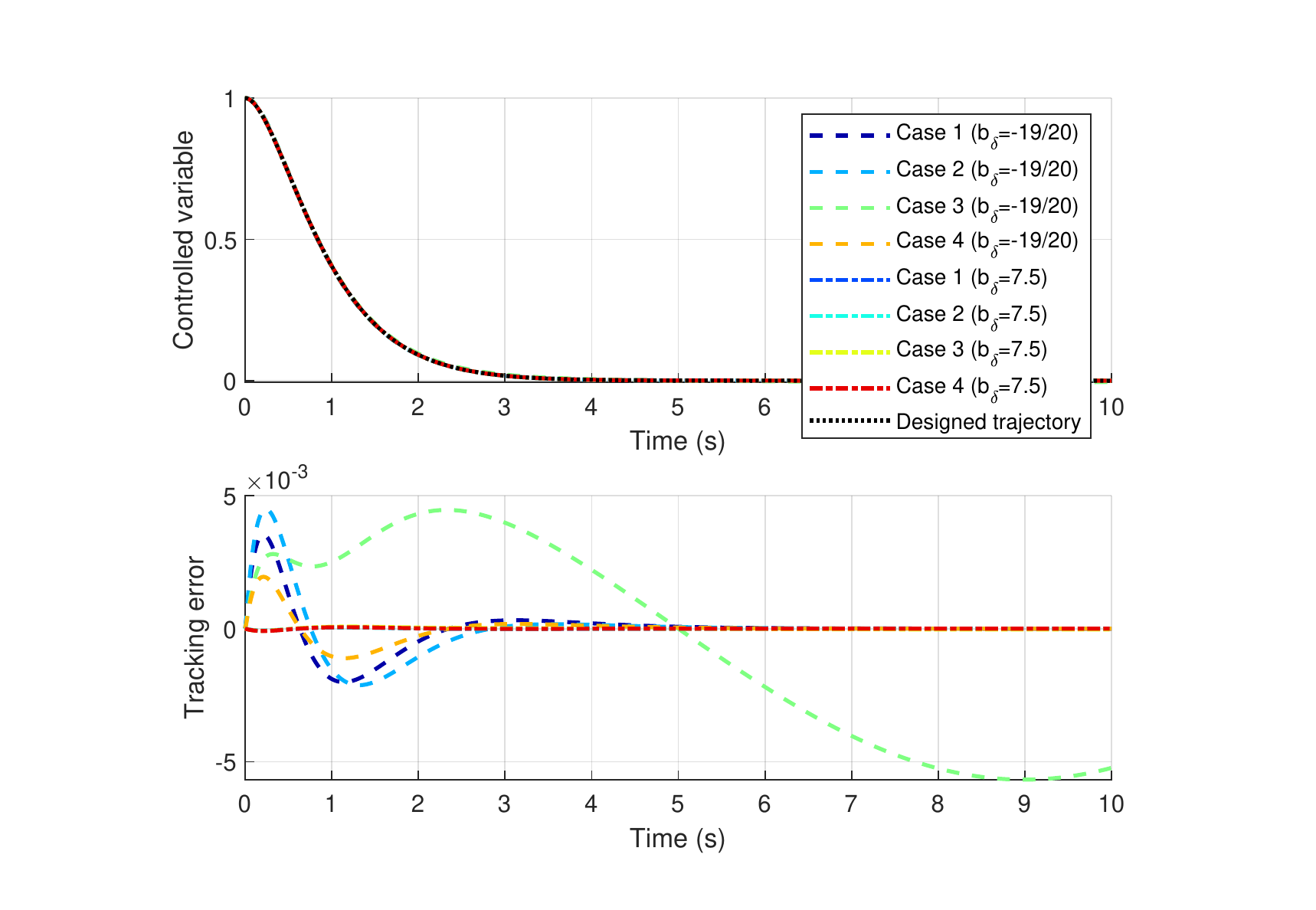}
      \caption{The tracking performance for the different uncertainties (\ref{DifferentUncertainty}) with $\phi_i$ in (\ref{SimuPhi1}).}
        \label{Figp1x1}
\end{figure}

\begin{figure}[t]\centering
      \includegraphics[width=1\linewidth,trim=50 250 50 250,clip]{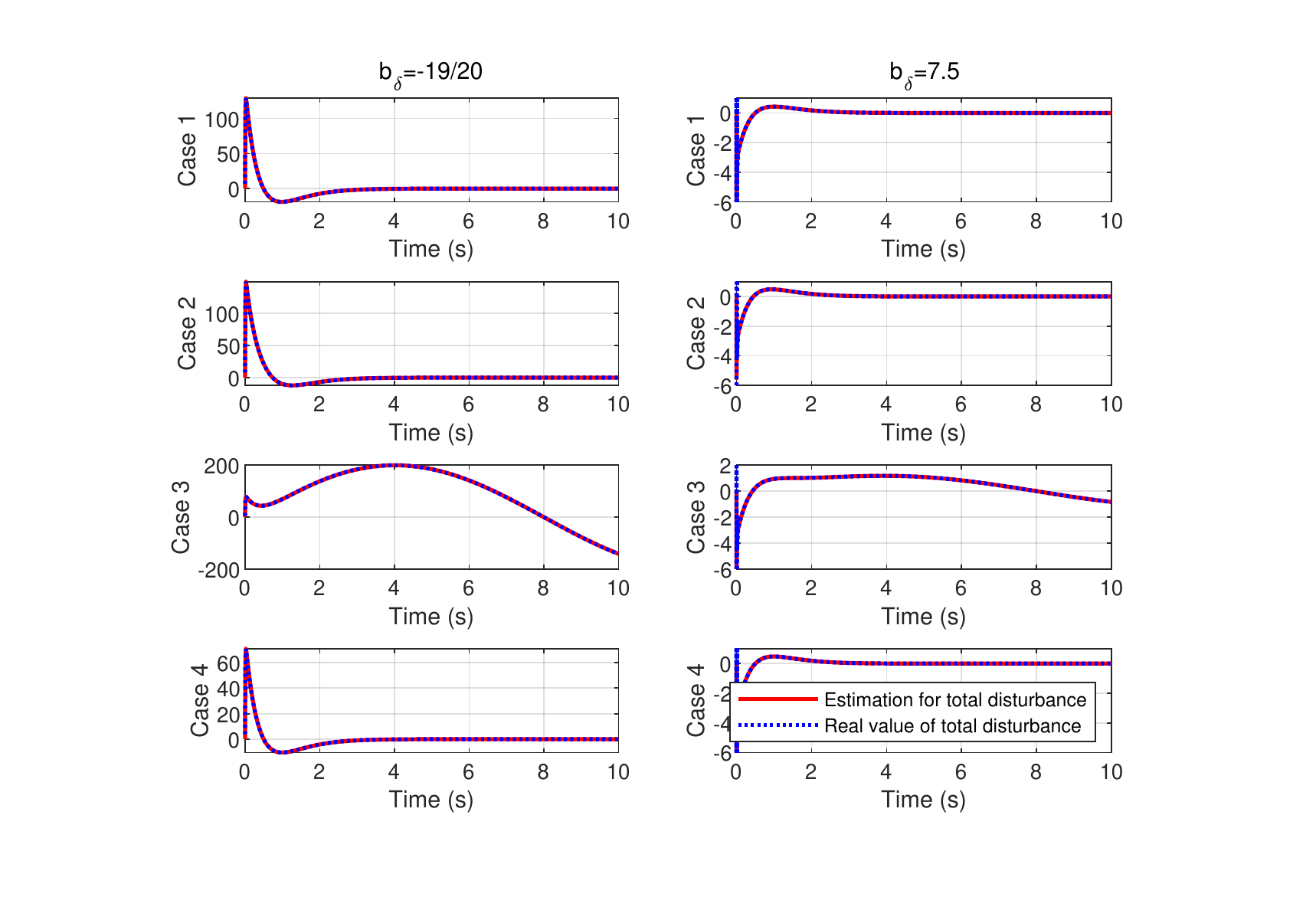}
      \caption{The estimation performance for the different uncertainties (\ref{DifferentUncertainty}) with $\phi_i$ in (\ref{SimuPhi1}).}
        \label{Figp1est3}
\end{figure}

Choose $\phi_i~(i=1,2,3)$ by the design method (\ref{bandwidthPHI}), i.e.,
\begin{equation}\label{SimuPhi1}
\phi_1 = 3,\quad \phi_2 = 3,\quad \phi_3 = 1.
\end{equation}
From Theorem~\ref{ThmStable}, the necessary and sufficient condition of the uncertain control gain is $b_\delta \in (-1,8)$.
Considering $b_\delta=-\frac{19}{20}$ and $b_\delta=7.5$, the simulation results for the four cases of uncertainties (\ref{DifferentUncertainty}) are shown in Fig.~\ref{Figp1x1}--\ref{Figp1est3}.
Fig.~\ref{Figp1x1} shows that, despite the various uncertainties including nonlinear dynamics and uncertain control gain, the tracking performance is highly consistent with the desired system via the ADRC controller (\ref{RESO})--(\ref{a1p1}).
From Fig.~\ref{Figp1est3}, the ESO (\ref{RESO}) is shown to effectively estimate the total disturbance, which contributes to the satisfactory performance of the closed-loop system.

%
%\begin{figure}[t]\centering
%      \includegraphics[width=1\linewidth,trim=50 250 50 250,clip]{est1P1.pdf}
%      \caption{The estimation performance for $b_\delta=-\frac{19}{20}$ and the different uncertainties (\ref{DifferentUncertainty}) with $\phi_i$ in (\ref{SimuPhi1}).}
%        \label{Figp1est1}
%\end{figure}
%
%\begin{figure}[t]\centering
%      \includegraphics[width=1\linewidth,trim=50 250 50 250,clip]{est2P1.pdf}
%      \caption{The estimation performance for $b_\delta=7.5$ and the different uncertainties (\ref{DifferentUncertainty}) with $\phi_i$ in (\ref{SimuPhi1}).}
%        \label{Figp1est2}
%\end{figure}

\begin{figure}[t]\centering
      \includegraphics[width=1\linewidth,trim=50 250 50 250,clip]{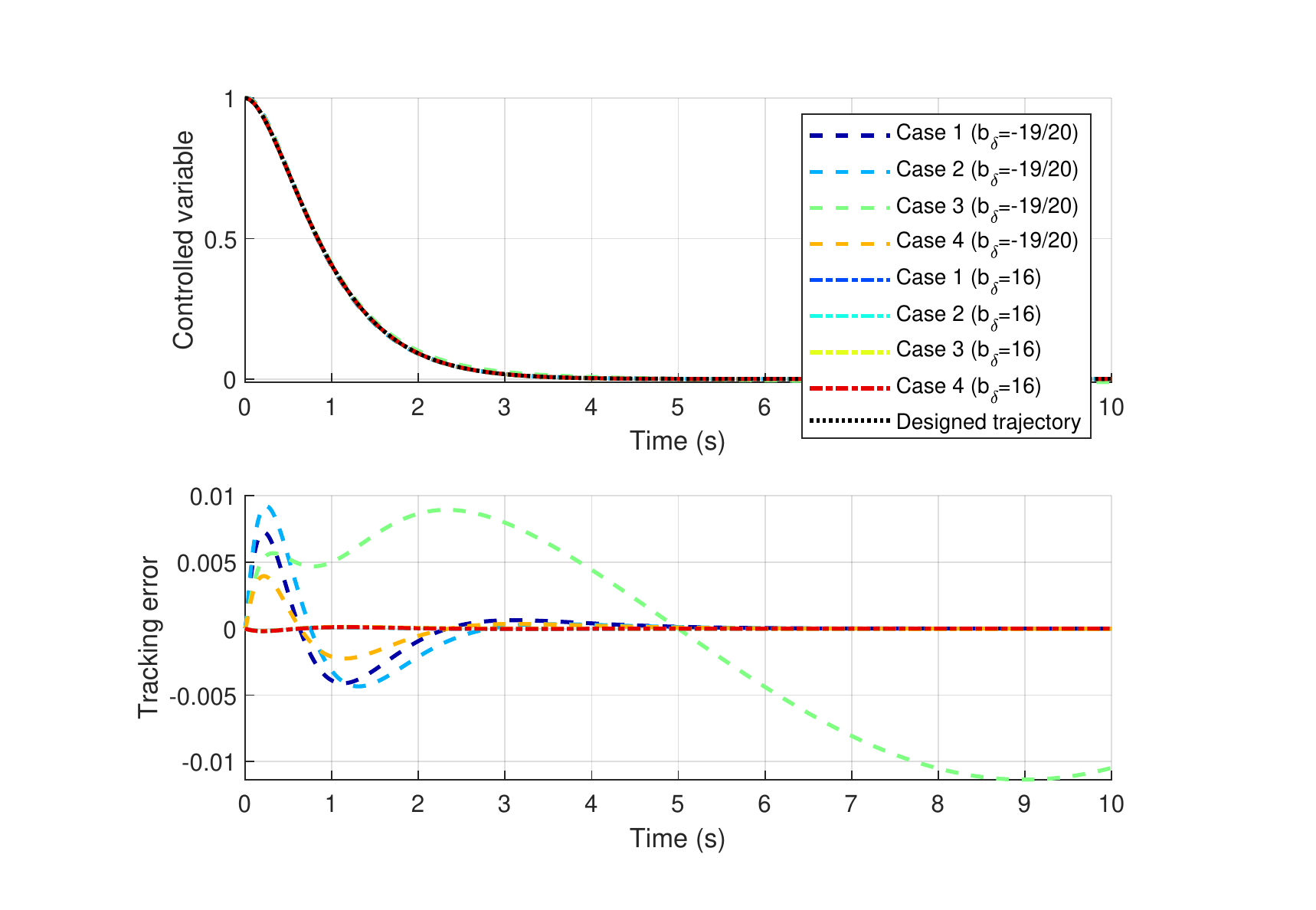}
      \caption{The tracking performance for the different uncertainties (\ref{DifferentUncertainty}) with $\phi_i$ in (\ref{SimuPhi2}).}
      \label{Figp05x1}
\end{figure}

From Theorem~\ref{ThmStable}, with the design of $\phi_i$ (\ref{SimuPhi1}), the closed-loop system will be unstable if the uncertain control gain $b_\delta$ is larger than 8.
One tuning law of $\phi_i$ to enlarge the scope of uncertain control gain is to select a smaller $\phi_3$, as shown in the following case.
Choosing
\begin{equation}\label{SimuPhi2}
\phi_1 = 3,\quad  \phi_2 = 3,\quad \phi_3 = 0.5,
\end{equation}
Theorem~\ref{ThmStable} implies that the necessary and sufficient condition of the uncertain control gain is that $b_\delta \in (-1,17)$.
Considering $b_\delta=-\frac{19}{20}$ and $b_\delta=16$, the simulation results for the $\phi_i$ satisfying (\ref{SimuPhi2}) and the four cases of uncertainties (\ref{DifferentUncertainty}) are shown in Fig.~\ref{Figp05x1}.
Fig.~\ref{Figp05x1} demonstrates that the tracking performance is highly consistent with the desired trajectory under the large uncertainty of control input.

\section{Conclusion}
The paper rigorously studies the capability of the ADRC to handle the uncertainties related with control input.
The necessary and sufficient condition for the well-performed ADRC design is rigorously studied, which further reveals the capability of handling the uncertain control gain.
The maximum range of the uncertain control gain is quantitatively presented, which greatly improves the existing results.
Moreover, under the proposed necessary and sufficient condition, both the transient performance and the steady state property of the ADRC based closed-loop system are rigorously analyzed.
%Furthermore, a tuning law of the ESO's parameters to extend the maximum range of the uncertain control gain is proposed and tested by simulations.

\section{Appendix}

{\bf Proof of Theorem~\ref{ThmStable}.}
Define the tracking error and estimating error as
$E(t)=X(t)-R(t)$ and $\xi =\begin{bmatrix}\xi_1~ \cdots ~ \xi_{n+1} \end{bmatrix}^T= T_1^{-1} \left[ \begin{smallmatrix} X(t)-\hat{X}(t) \\ f(X,u,t) - \hat{f}(t) \end{smallmatrix} \right]$ where $T_1 =\left[ \begin{smallmatrix} \omega_o^{-n} &0& \cdots &0\\0&\omega_o^{1-n}&\ddots&\vdots\\ \vdots&\ddots&\ddots&0\\0&\cdots&0&1 \end{smallmatrix} \right]$.
Then, the control input (\ref{control_ADRC})
can be reformulated as follows:
\begin{equation}\label{proof_t3_u2}
u(t) =\left\{\begin{split}
&0,\quad t_0 \leq t< t_u,\\
&\frac{-g(E+T_1 \xi,t)-K^T E(t)+K_e^T T_1 \xi(t)+r^{(n)}(t) }{b},\\ &\qquad\qquad\qquad\qquad\qquad\qquad\qquad t\geq t_u,
\end{split}\right.
\end{equation}
Therefore, the closed-loop system (\ref{plant}) and (\ref{RESO})--(\ref{a1p1})
can be written as follows:
\begin{equation}\label{proof_t3_exi_1}
\left\{ \begin{split}
&\dot{E} = A E+B \Gamma_0(E,t),\\
&\dot{\xi}=\omega_o A_{1}\xi+B_2 \Gamma_1(E,t),
\end{split}\right.
\quad t_0 \leq t< t_u,
\end{equation}
\begin{equation}\label{proof_t3_exi_2}
\left\{ \begin{split}
&\dot{E} = A_K E + B K_{e}^T T_1 \xi,\\
&\dot{\xi}=\omega_o A_{2}\xi+B_2 \Gamma_2(E,u,\xi,t) ,
\end{split}\right.
\quad t\geq t_u,
\end{equation}
where $B_2 = \begin{bmatrix} 0 \\ B \end{bmatrix}$, $ K_{e} = \begin{bmatrix} K \\ 1\end{bmatrix}$,
$\Gamma_0(E,t) = f(E+R,0,t)-r^{(n)}$,
$\Gamma_1(E,t)=\frac{\partial f(E+R,0,t)}{\partial t}+\frac{\partial f(E+R,0,t)}{\partial (E+R)}(A(E+R)+Bf(E+R,0,t) )$,
$\Gamma_2(E,u,\xi,t) = \frac{\partial f(E+R,u,t)}{\partial (E+R)}(A_KE+AR+Br^{(n)}) +\frac{\partial f(E+R,u,t)}{\partial t}+({b_\delta}/ \bar{b})(-K^T A_KE+r^{(n+1)})+  \frac{\partial f(E+R,u,t)}{\partial (E+R)} B K_e^T T_1 \xi + ({b_\delta}/ \bar{b}) (K^T T_2 -K^T B [K^T~0] T_1)  \xi $,
and
$T_2 =\left[\begin{smallmatrix} -\phi_1\omega_o^{1-n} & \omega_o^{1-n} & 0 &\cdots &0 &0 \\ -\phi_2 \omega_o^{2-n} &0 & \omega_o^{2-n}&\ddots & \vdots & \vdots \\ \vdots & \vdots & \ddots & \ddots &0 &\vdots \\ -\phi_{n-1} \omega_o^{-1}&\vdots &\ddots &\ddots& \omega_o^{-1} &0 \\ -\phi_n & 0 &\cdots &\cdots &0 &0 \end{smallmatrix}\right]$.

\textbf{Sufficiency~$(S1\Rightarrow S2)$:}
Firstly, the properties of $A_K$, $A_1$, $A_2$ and $\Gamma_i~(0\leq i\leq 2)$ in the closed-loop system (\ref{proof_t3_exi_1})-(\ref{proof_t3_exi_2}) are analyzed, respectively.

Since $A_K$, $A_1$ and $A_2$ are Hurwitz, there exist positive definite matrices $P_K$, $P_1$ and $P_2$ such that
\begin{equation}\label{proof_t3_13}
A_K^T P_K+P_K A_K=-\emph{\textbf{I}},~ A_1^T P_1+P_1A_1=-\emph{\textbf{I}},~
A_2^T P_2+P_2A_2 = -\emph{\textbf{I}}.
\end{equation}

Define the function $\Psi(a) \triangleq \sup_{\|x\|\leq a} \psi(x)$,
which is non-decreasing with respect to the variable $a$.
Let $\rho_e$ be any positive. Then, $\Gamma_i~(i=0,1)$ has the following bound:
\begin{equation}\label{proof_t3_10}
|\Gamma_0| \leq \pi_0(\rho_e),\quad
|\Gamma_1| \leq \pi_1(\rho_e),\quad
\end{equation}
where the positives $\pi_0(\rho_e) \triangleq \Psi(\rho_e)+\eta_r$ and $\pi_1(\rho_e) \triangleq \Psi(\rho_e)(1+\|A\|(\rho_e+\sqrt{n}\eta_r)+\Psi(\rho_e))$
are non-decreasing with respect to the variable $\rho_e$.

Let $\rho_e$, $\rho_\xi$ and $\omega_0^*$ be any positives. Then, for any $E\in\{E|~\|E\|\leq \rho_e\}$, $\xi\in\{\xi|~\|\xi\|\leq \rho_\xi\}$ and $\omega_o\in\{\omega_o|~\omega_o\geq \omega_0^*\}$, it can be deduced from (\ref{proof_t3_u2}) that
\begin{equation}\label{proof_t3_11_2}
\begin{split}
|u(t)| &\leq \eta_u (\rho_e,\rho_\xi,\omega_0^*) ,
\end{split}
\end{equation}
where $\eta_u (\rho_e,\rho_\xi,\omega_0^*) = (\Psi(\rho_e+\|T_1(\omega_0^*)\| \rho_\xi)+\eta_r+\|K\|\rho_e+ \|K_e^T\|\|T_1(\omega_0^*)\| \rho_\xi)/b$.
From Assumption~\ref{ass_uncertainty}, the bound of the total disturbance and its partial derivatives can be expressed as $|f(E+R,u,t)|\leq \Psi_{f,1}(\rho_e,\rho_\xi,\omega_0^*)$, $\left\|\frac{\partial f(E+R,u,t)}{\partial (E+R)}\right\|\leq \Psi_{f,2}(\rho_e,\rho_\xi,\omega_0^*)$ and $\left|\frac{\partial f(E+R,u,t)}{\partial t}\right|
\leq \Psi_{f,2}(\rho_e,\rho_\xi,\omega_0^*)$,
where $\Psi_{f,1}(\rho_e,\rho_\xi,\omega_0^*)\triangleq \Psi(\rho_e+\sqrt{n}\eta_r)+b_\delta \eta_u (\rho_e,\rho_\xi,\omega_0^*)$
and
$\Psi_{f,2}(\rho_e,\rho_\xi,\omega_0^*)\triangleq \Psi(\rho_e+\sqrt{n}\eta_r)$.
Therefore, $\Gamma_2$ has the following bound:
\begin{equation}\label{proof_t3_14}
|\Gamma_2| \leq \pi_2(\rho_e,\rho_\xi,\omega_0^*),
\end{equation}
where the positive $\pi_2(\rho_e,\rho_\xi,\omega_0^*) \triangleq \left|\frac{b_\delta}{b}\right| \cdot (\|K^T A_K\| \rho_e + \eta_r + (\|K^T T_2(\omega_0^*)\| +\|K^T B [K^T~0]\|\| T_1(\omega_0^*)\| )\rho_\xi ) + \Psi_{f,2}(\rho_e, \eta_u (\rho_e,\rho_\xi,\omega_0^*)) (1+\|A_K\|\rho_e + (1+\sqrt{n}\|A\|)\eta_r + \|K_e^T\|\| T_1(\omega_0^*)\|\rho_\xi )$
are non-decreasing with respect to $(\rho_e,\rho_\xi)$ and non-increasing with respect to $\omega_0^*$.

Then, with the similar technique in \cite{xue2018performance}, the trajectories of the tracking error and the estimation error in the time sequences $[t_0,t_u)$ and $[t_u,\infty)$ can be analyzed, which implies that $(S1 \Rightarrow S2)$.
The details is shown in the supplementary file.

\textbf{Necessity~$(S2\Rightarrow S1)$:}
The proof is based on reduction to absurdity.
Assume that for any given $\varepsilon$, there exists $\omega_o$ such that (\ref{ThmStableX*}) is satisfied and the matrix $A_2$ is not Hurwitz.

Let $g(X,t) = 1$ and $r=0$. From (\ref{plant}) and (\ref{ThmStableX*}), it can be deduced that
\begin{equation}\label{bneq0}
\bar{b}+b_\delta \neq 0.
\end{equation}

Let $g(X,t) = M_g\sin(w_g t+\phi_g)$, $r=0$ and $X(t_0) = \emph{\textbf{0}}$. Combined with (\ref{idealTra}), we have $X^*(t) \equiv \emph{\textbf{0}}$.
Moreover, (\ref{proof_t3_exi_2}) can be reformulated as follows.
\begin{equation}\label{caseXi}
\dot{\xi}=A_3 \xi +B_2(-\frac{b_\delta}{\bar{b}} K^T A_k E+M_g w_g \cos(w_g t+\phi_g)),
\end{equation}
where $A_3 = \omega_o A_2 + B_2 (K^T T_2 -K^T B \begin{bmatrix}K^T&0 \end{bmatrix}T_1) \frac{b_\delta}{b}$.
%\begin{equation}\label{newA3}
%A_3 = \omega_o A_2 + B_2 (K^T T_2 -K^T B \begin{bmatrix}K^T&0 \end{bmatrix}T_1) \frac{b_\delta}{b}.
%\end{equation}

Denoting $\hat{E} (t)\triangleq \begin{bmatrix}\hat{e}_1 (t)& \cdots & \hat{e}_{n+1}(t) \end{bmatrix}^T = T_1 \xi$, $E_z(t) \triangleq E(t) - e^{A_{k_c}(t-t_0)}E(t_0) $,
the dynamics of $\hat{E}$ and $E_z$ are shown as follows:
\begin{equation}\label{DynamicEzHatE}
\left\{ \begin{split}
&\dot{E}_z = A_K{E}_z + B K_e^T \hat{E},\\
&\dot{\hat{E}} = \bar{A}_3 \hat{E} + B_2 (\Gamma_3+M_g w_g \cos(w_g t+\phi_g)),
\end{split} \right.
~
t\in[t_u,\infty),
\end{equation}
where
\begin{equation}
\left\{ \begin{split}
&\Gamma_3 = -\frac{b_\delta}{\bar{b}} K^T A_k E = -\frac{b_\delta}{\bar{b}} K^T A_k E_z, \\
&\bar{A}_3 = T_1 A_3 T_1^{-1}
=\left[\begin{smallmatrix}-\omega_o\phi_1& 1&\cdots& 0 \\ -\omega_o^2 \phi_2 & 0 & \ddots &  \vdots \\   \vdots & \vdots & \ddots &   1 \\ -\omega_o^{n+1}\phi_{n+1} \left(1+ \frac{b_\delta}{\bar{b}}\right)+\tilde{a}_1 & \cdots  &\tilde{a}_n & 0 \end{smallmatrix}\right],\\
&\tilde{a}_1 = k_n k_1-\Sigma_{i=1}^{n} \phi_i k_i \omega_o^i,\\
&\tilde{a}_j = k_{j-1} - k_n k_j,\quad 2\leq j\leq n.
\end{split} \right.
\end{equation}

According to (\ref{DynamicEzHatE}), there is $\dot{x}_{n}-\dot{x}_{n}^* = -\Sigma_{i=1}^{n} k_i (x_{i}-x_{i}^*)  + K_e^T \hat{E}$.
Since (\ref{ThmStableX*}) is satisfied, it can be deduced that
\begin{equation}\label{bounddotxn}
-n \varepsilon \max_{1\leq i\leq n}\{k_i\}  + K_e^T \hat{E} \leq \dot{x}_{n}-\dot{x}_{n}^* \leq n \varepsilon \max_{1\leq i\leq n}{k_i}  + K_e^T \hat{E},
\end{equation}
By reduction to absurdity, we will prove that for any given $T_0>0$ and $T\geq t_u$,
\begin{equation}\label{InfBoundKHatE}
\inf_{t\in[T,T+T_0]}|K_e^T \hat{E}| \leq \eta_\varepsilon,
\end{equation}
where $\eta_\varepsilon = \frac{3\varepsilon+ T_0 n \varepsilon \max_{1\leq i\leq n} \{k_i\}}{T_0}$.
Assume that there exists $T_0>0$ and $T\geq t_u$ such that
\begin{equation}\label{fzfAssume}
\inf_{t\in[T,T+T_0]} K_e^T \hat{E}  > \eta_\varepsilon.
\end{equation}
According to (\ref{ThmStableX*}), it can be obtained that $|x_n(T)-x_n^*(T)|\leq \varepsilon$.
From (\ref{bounddotxn}) and (\ref{fzfAssume}), it can be calculated that
\begin{equation}\nonumber
\begin{split}
&\qquad x_n(T+T_0)-x_n^*(T+T_0) \\
&\geq x_n(T)-x_n^*(T) + \left( -n \varepsilon \max_{1\leq i\leq n}\{k_i\}  + \inf_{t\in[T,T+T_0]} K_e^T \hat{E} \right) T_0\\
& \geq  -\varepsilon +  3\varepsilon = 2\varepsilon,
\end{split}
\end{equation}
which contradicts to (\ref{ThmStableX*}).
Thus, for any $T_0>0$ and $T\geq t_u$, we have
\begin{equation}\label{pro_fzf2}
\inf_{t\in[T,T+T_0]} K_e^T \hat{E}  \leq \eta_\varepsilon.
\end{equation}
Similar with the deduction (\ref{fzfAssume})--(\ref{pro_fzf2}), it can be deduced that
\begin{equation}\label{pro_fzf3}
\inf_{t\in[T,T+T_0]} K_e^T \hat{E}  \geq  -\eta_\varepsilon.
\end{equation}
From (\ref{pro_fzf2}) and (\ref{pro_fzf3}), (\ref{InfBoundKHatE}) is proved.

Since the amplitude $M_g$, the angular velocity $w_g$ and the phase angle $\phi_g$ in (\ref{DynamicEzHatE}) can be arbitrarily chosen and $B_2 \Gamma_3$ is bounded, as shown in the following equation,
\begin{equation}\label{boundGamma3}
\| B_2 \Gamma_3 \| \leq \left| \frac{b_\delta}{\bar{b}} \right| \|K^T A_K\| \varepsilon,
\end{equation}
the necessary condition for (\ref{InfBoundKHatE}) is that the matrix $\bar{A}_3$ is Hurwitz. Since the eigenvalues of $A_3$ and $\bar{A}_3$ are the same, the necessary condition for (\ref{InfBoundKHatE}) is that $\omega_o\in \Omega_{\omega_o}$
where $\Omega_{\omega_o} \triangleq \{ \omega_o>0 ~|~ A_3 \text{~is~Hurwitz}   \}$.
Since (\ref{bneq0}) is hold, the form of $A_3$ implies that
\begin{equation}\label{limitA3}
\lim_{\omega_o \rightarrow \infty} \frac{\|A_3 - \omega_o A_2 \|}{\omega_o}=0.
\end{equation}
Since $A_2$ is not Hurwitz, (\ref{limitA3}) implies that there exists $\bar{\omega}_o>0$ such that $A_3$ is not Hurwitz for $\omega_o \geq \bar{\omega}_o$.
Thus, $\Omega_{\omega_o}\subseteq (0,\bar{\omega}_o)$ is a bounded set.

Next, we will prove that there exists a uncertainty $M_g \sin(\omega_g t +\phi_g)$ such that (\ref{ThmStableX*}) is not satisfied, if $A_3$ is Hurwitz, i.e., $\Omega_{\omega_o}\subseteq (0,\bar{\omega}_o)$.

Firstly, we will prove that for any given $M_e>0$, $T_0>0$, $T>t_u$, and $\omega_o \in (0,\bar{\omega}_o)$, there exists a uncertainty $M_g \sin(\omega_g t +\phi_g)$ such that
\begin{equation}\label{keypro1}
\inf_{t\in[T+T_0,T+2T_0]} \|\hat{E}(t)\| \geq M_e.
\end{equation}
From (\ref{DynamicEzHatE}), the trajectory of $\hat{E}$ satisfies that $\hat{E}(t) = e^{\bar{A}_3 (t-T)} \hat{E}(T) + \int_T^t e^{\bar{A}_3 (t-s) } B_2 (\Gamma_3(s)+M_g w_g \cos(w_g s+\phi_g))  d s$.
By selecting $(M_g,\omega_g,\phi_g)=(M_{g,1},\omega_{g,1},\phi_{g,1})$ and $M_{g,1}>0$ which satisfies $\inf_{s\in[T,T+2T_0]} (M_{g,1} w_{g,1} \cos(w_{g,1} s+\phi_{g,1}))>0$,
the bound of $\hat{E}$ can be obtained as follows.
\begin{equation}\label{boundHatE}
\begin{split}
&\| \hat{E}(t) \| \geq \inf_{s\in[T,t]} (M_g w_g \cos(w_g s+\phi_g)) \cdot \|\int_T^t e^{\bar{A}_3 (t-s) } B_2  d s \| \\
&~~ - \left| \frac{b_\delta}{\bar{b}} \right| \|K^T A_K\| \varepsilon \cdot  \int_T^t \| e^{\bar{A}_3 (t-s) } B_2 \| d s - \|e^{\bar{A}_3 (t-T)}\|  \| \hat{E}(T) \|
\end{split}
\end{equation}
for $t\in[T,T+2T_0]$.

Denote $M_1(T_0) = \inf_{m\in[T_0,2T_0],\omega_o\in\Omega_{\omega_o}} \| \int_0^m e^{\bar{A}_3 (m-\tau)} B_2 d\tau \|$, $M_2(T_0) = \sup_{m\in[T_0,2T_0],\omega_o\in\Omega_{\omega_o}}  \int_0^m \|e^{\bar{A}_3 (m-\tau)} B_2 \| d\tau$ and $M_3(T_0) =\sup_{m\in[T_0,2T_0],\omega_o\in\Omega_{\omega_o}} \|e^{\bar{A}_3 (m-\tau)}\|$.
Since the trace of $\bar{A}_3$ is $-\omega_o \phi_1$ which equals the sum of all the eigenvalues of $\bar{A}_3$, we have
\begin{equation}\label{LowBoundReA3}
\min_{\omega_o\in\Omega_{\omega_o},\sigma \in \lambda (\bar{A}_3)} Re(\sigma) \geq - \bar{\omega}_o \phi_1.
\end{equation}
Since $\bar{A}_3$ is Hurwitz, there is
\begin{equation}\label{UpperBoundReA3}
\max_{\omega_o\in\Omega_{\omega_o},\sigma \in \lambda (\bar{A}_3)} Re(\sigma) \leq 0.
\end{equation}
From (\ref{LowBoundReA3}) and (\ref{UpperBoundReA3}), $M_1(T_0)$ has a positive lower bound and $M_2(T_0)$ and $M_3(T_0)$ have a positive upper bounded for any given $T_0>0$, i.e., $M_1(T_0) \geq \b{M}_1(T_0)>0$, $M_2(T_0) \leq \bar{M}_2(T_0)$ and $M_3(T_0) \leq \bar{M}_3(T_0)$, where $(\b{M}_1(T_0),\bar{M}_2(T_0),\bar{M}_3(T_0))$ are positive constants.

Then, selecting $(\omega_g,\phi_g) = (\omega_{g,1},\phi_{g,1})$, there exists $M_{g,2}\geq M_{g,1}>0$ such that for $M_g\geq M_{g_2}$,
\begin{equation}\label{ChooseMgOgPg}
\begin{split}
&\inf_{s\in[T,T+2T_0]} (M_{g} w_{g} \cos(w_{g} s+\phi_{g})) \\
&\qquad > \frac{M_e+\bar{M}_3(T_0)\|\hat{E}(T)\|+\left| \frac{b_\delta}{\bar{b}} \right| \|K^T A_K\| \varepsilon\bar{M}_2(T_0)}{\b{M}_1(T_0)}.
\end{split}
\end{equation}
From (\ref{boundHatE})--(\ref{ChooseMgOgPg}), the bound of $\hat{E}$ satisfies the following equation under the uncertainty $M_{g,2} \sin(\omega_{g,1} t +\phi_{g,1})~(M_g\geq M_{g_2})$ and $\omega_o\in \Omega_{\omega_o}$.
\begin{equation}
\begin{split}
&~~~~\inf_{t\in[T+T_0,T+2T_0]} \| \hat{E}(t) \| \\
&\geq \inf_{s\in[T,T+2T_0]} (M_g w_g \cos(w_g s+\phi_g)) \b{M}_1(T_0) \\
&~~~~- \left| \frac{b_\delta}{\bar{b}} \right| \|K^T A_K\| \varepsilon \bar{M}_2(T_0) - \bar{M}_3(T_0) \|\hat{E}_T\|\\
&\geq M_e.
\end{split}
\end{equation}
Thus, (\ref{keypro1}) is proved.

Denote $\tilde{E} =[\tilde{e}_1~\cdots ~\tilde{e}_{n+1} ]^T$ and consider the following differential equation.
\begin{equation}\label{FuzhuSys}
\dot{\tilde{E}} = \bar{A}_3 \tilde{E} + B_2 M,
\end{equation}
where $M>0$ is a constant.
Due to the specific form of $\bar{A}_3$ and $B_2$, for the given initial value $\tilde{E}(T)=\hat{E}(T)$, there exists a sufficiently large $M^*$ such that $\inf_{t\in[T+T_0,T+2T_0]} \tilde{e}_i(t)>0$ for $1\leq i \leq n+1$, $M\geq M^*$ and $\omega_o\in\Omega_{\omega_o}$.

There exists $M_g^*\geq M_{g,2}$ such that
\begin{equation}\label{ChooseMgOgPg22}
\inf_{s\in[T,T+2T_0]} (M_{g} w_{g} \cos(w_{g} s+\phi_{g})) \geq M^* +\left| \frac{b_\delta}{\bar{b}} \right| \|K^T A_K\| \varepsilon ,
\end{equation}
for $(\omega_g,\phi_g) = (\omega_{g,1},\phi_{g,1})$ and $M_g\geq M_g^*$.
Combined with the boundary of $\Gamma_3$ shown in (\ref{boundGamma3}), we have
\begin{equation}\label{boundMgcos}
\inf_{s\in[T,T+2T_0]} (\Gamma_3(s)+ M_{g} w_{g} \cos(w_{g} s+\phi_{g})) \geq M^*.
\end{equation}
From (\ref{FuzhuSys})--(\ref{boundMgcos}), it can be concluded that
the components of $\hat{E}$ satisfy
\begin{equation}\label{keypro2}
\inf_{t\in[T+T_0,T+2T_0]} \hat{e}_i(t)>0, \quad 1\leq i \leq n+1,
\end{equation}
for the uncertainty $M_{g} \sin(\omega_{g,1} t +\phi_{g,1})~(M_g\geq M_{g}^*)$ and $\omega_o\in \Omega_{\omega_o}$.

With the combination of (\ref{keypro1}) and (\ref{keypro2}), there is
\begin{equation}\label{keypro3}
\begin{split}
\inf_{t\in[T+T_0,T+2T_0]} |K_e^T \hat{E}(t)| \geq \min_{1\leq i\leq n} \{k_i,1\} \cdot M_e.
\end{split}
\end{equation}
By choosing $M_e = \frac{2\eta_\varepsilon}{\min_{1\leq i\leq n} \{k_i,1\}}$, (\ref{keypro3}) contradicts to (\ref{InfBoundKHatE}), which implies that $A_2$ should be Hurwitz.
\hfill $\square$

%%%%%%%%%%%%%%%%%%Thm2
{\bf Proof of Theorem~\ref{Thm_converge}:}
Due to the special form of $g(X,t)$ as shown in Assumption~\ref{ass_limit}, there is $\frac{\partial f}{\partial X} = \frac{\partial g_1}{\partial X}$ and $\frac{\partial f}{\partial t} = \dot{d}_1(t)$.
Moreover, the limitation of the external disturbance exists, i.e.,
$\lim_{t\rightarrow \infty} d_1(t) = \bar{D}$,
where $\bar{D}$ is a constant.

Recall the definition of $(E,\xi)$ and define $\bar{\xi} \triangleq \xi - B_2 (d_1(t)-\bar{D})$
Then, the ADRC based closed-loop system for $t\in[t_u,\infty)$ (\ref{proof_t3_exi_2}) can be rewritten as
\begin{equation}\label{proD_6}
\dot{E} = A_{K} E + B K_{e}^T T_1 \bar{\xi}+\bar{\Gamma}_0,~
\dot{\bar{\xi}}=\omega_o A_{2}\bar{\xi}+\bar{\Gamma}_1+\bar{\Gamma}_2,~ t\geq t_u,
\end{equation}
where $\bar{\Gamma}_0 = B K_{e}^T T_1 B_2(d_1(t)-\bar{D})$,
$\bar{\Gamma}_1 =  B_2\frac{\partial g_1(E+R)}{\partial (E+R)} \left[\begin{smallmatrix} r^{(1)} & \cdots & r^{(n)} \end{smallmatrix}\right]^T+B_2 \frac{b_\delta}{\bar{b}} r^{(n+1)}+ A_2B_2\omega_o(d_1(t)-\bar{D}) - B_2 \frac{\partial g_1(E+R)}{\partial (E+R)} B K_e^T T_1
(d_1(t)-\bar{D}) - B_2 \frac{b_\delta}{\bar{b}} (K^T T_2 -K^T B [K^T~0] T_1)
(d_1(t)-\bar{D})$
and
$\bar{\Gamma}_2 =B_2\left( \frac{\partial g_1(E+R)}{\partial (E+R)} A_K-\frac{b_\delta}{\bar{b}} K^T A_K\right) E\quad + B_2 \frac{\partial g_1(E+R)}{\partial (E+R)} B K_e^T T_1 \xi  + B_2 \frac{b_\delta}{\bar{b}} (K^T T_2 -K^T B \begin{bmatrix}K^T&0 \end{bmatrix}T_1) \xi$.

Since all assumptions in Remark~\ref{Prop_transient} are satisfied, according to (\ref{Thm_re1_track})--(\ref{Thm_re2_est}), there exists positives $\eta_{E}^*$, $\eta_{\Xi}^*$ and $\omega_o^*$ such that
\begin{equation}\label{proD_8}
\sup_{t\geq t_0}|E(t)|\leq \eta_{E}^*,\quad
\sup_{t\geq t_0}|\xi(t)|\leq \eta_{\Xi}^*,
\end{equation}
for any $\omega_o\geq \omega_o^*$.

From (\ref{proD_8}), the bounds of $\bar{\Gamma}_0$,  $\bar{\Gamma}_1$ and $\bar{\Gamma}_2$ for $\omega_o\geq \omega_o^*$ are shown as follows.
\begin{equation}\label{proD_11}
\left\{\begin{split}
\|\bar{\Gamma}_0\| \leq & \gamma_{01} |d_1(t)-\bar{D}|,\\
\|\bar{\Gamma}_1\| \leq & \gamma_{11} |d_1(t)-\bar{D}|+ \gamma_{12} \left\|\left[\begin{smallmatrix} r^{(1)}(t) & r^{(2)}(t) & \cdots r^{(n+1)}(t) \end{smallmatrix}\right] \right\| ,\\
\|\bar{\Gamma}_2\| \leq &  \gamma_{21}\|E\|+\gamma_{22}\|\bar{\xi}\|,
\end{split}\right.
\end{equation}
where $\gamma_{01} = \| BK_e^T\|\|T_1(\omega_o^*)\|\|B_2\|$, $\gamma_{11} = \omega_o+\gamma_{22}$, $\gamma_{12} = \sqrt{(\Psi(\eta_{E}^*+\sqrt{n}\eta_r))^2+  \left| {b_\delta}/{\bar{b}} \right|^2}$, $\gamma_{21} =\Psi(\eta_{E}^*+\sqrt{n}\eta_r)\|A_K\| +\left| \frac{b_\delta}{\bar{b}} \right| \|K^T A_K\|$ and $\gamma_{22} = \Psi(\eta_{E}^*+\sqrt{n}\eta_r) \|BK_e^T\|\| T_1(\omega_o^*) \|+ \left| \frac{b_\delta}{\bar{b}} \right| (\|K^T \|\| T_2(\omega_o^*)\| + \| K^T B  [K^T~0 ]\|\|T_1(\omega_o^*) \|) $.

Recall the definition of $P_K$ and $P_2$ (\ref{proof_t3_13}).
Denote
\begin{equation}\label{proD_12}
\bar{V} (E,\bar{\xi}) = E^T P_K E + \bar{\xi}^T P_2 \bar{\xi}.
\end{equation}
Combined with the definition of $c_{k1}$, $c_{k2}$, $c_{21}$, and $c_{22}$, (\ref{proD_12}) implies that
\begin{equation}\label{proD_12.5}
\left\{ \begin{split}
&\bar{V} (E,\bar{\xi}) \geq \min\{ c_{k1},c_{21} \} (\|E\|^2+\|\bar{\xi}\|^2),\\
&\bar{V} (E,\bar{\xi})
\leq \max\{ c_{k2},c_{22}\} (\|E\|^2+\|\bar{\xi}\|^2).
\end{split} \right.
\end{equation}
Based on (\ref{proD_6}), the derivative of $\sqrt{\bar{V}}$ satisfies the following bound: $\frac{d \sqrt{\bar{V}} }{d t}  \leq \frac{-\|E\|^2-\omega_o\|\bar{\xi}\|^2  }{2 \sqrt{\bar{V}}} +\frac{ 2 \|P_K BK_e^T\|\|T_1(\omega_o^*) \| \|E\|\|\bar{\xi}\|   }{2 \sqrt{\bar{V}}} +\frac{ 2  \|P_2\| \gamma_{21}\|E\|\|\bar{\xi}\| + 2 \|P_2\| \gamma_{22}\|\bar{\xi}\|^2 }{2 \sqrt{\bar{V}}} + \frac{ 2\|P_K\| \|E\| \|\bar{\Gamma}_0\| + 2 \|P_2\| \|\bar{\xi}\| \|\bar{\Gamma}_1\| }{2 \sqrt{\bar{V}}}$ for $\omega_o\geq \omega_o^*$.
By defining $\omega_o^{**} = \max\{ 2\left( 4\|P_K BK_e^T\|\|T_1(\omega_o^*) \|   + 4  \|P_2\| \gamma_{21} \right)^2+4\|P_2\|\gamma_{22} ,\omega_o^*\}$, we have $\frac{d \sqrt{\bar{V}} }{d t}  \leq \frac{-\frac{3}{4}\|E\|^2-\frac{\omega_o}{2}\|\bar{\xi}\|^2  }{2 \sqrt{\bar{V}}} + \frac{ 2\|P_K\| \|E\| \|\bar{\Gamma}_0\| + 2 \|P_2\| \|\bar{\xi}\| \|\bar{\Gamma}_1\| }{2 \sqrt{\bar{V}}}$
for $\omega_o\geq \omega_o^{**}$.
Combined with (\ref{proD_12.5}), there is $\frac{d \sqrt{\bar{V}} }{d t}  \leq -\bar{\alpha}_1 \sqrt{\bar{V}} + \bar{\alpha}_2 \sqrt{\|\bar{\Gamma}_0\|^2 + \|\bar{\Gamma}_1\|^2}$
where $\bar{\alpha}_1 = \frac{ \min\left\{\frac{3}{4} , \frac{\omega_o}{2}\right\} }{ 2\sqrt{\max\{c_{k2},c_{22}\}} }$ and $\bar{\alpha}_2 = \frac{\sqrt{c_{k2}^2+c_{22}^2}}{\min \{c_{k1},c_{21} \} }$.
By Grownwall Lemma, $\sqrt{\bar{V}(E,\bar{\xi})}$ has the following bound:
\begin{equation}\label{proD_15}
\begin{split}
\sqrt{\bar{V}(E,\bar{\xi})} \leq & \sqrt{\bar{V}(E(t_0),\bar{\xi}(t_0))} e^{-\bar{\alpha}_1(t-t_0)}\\
& +\int_{t_0}^{t} e^{-\bar{\alpha}_1(t-\tau)} \sqrt{\|\bar{\Gamma}_0(\tau)\|^2 + \|\bar{\Gamma}_1(\tau)\|^2} d\tau.
\end{split}
\end{equation}
Since $\lim_{t\rightarrow \infty} \|\bar{\Gamma}_0(t)\|  =0$ and $\lim_{t\rightarrow \infty} \|\bar{\Gamma}_1(t)\| =0$,
we have $\lim_{t\rightarrow \infty} \sqrt{\|\bar{\Gamma}_0(t)\|^2 + \|\bar{\Gamma}_1(t)\|^2} =0$.
According to Assumptions~\ref{ass_uncertainty}--\ref{ass_limit}, there is $|d_1(t)-d_1(t_0)| \leq |g_1(X)+d_1(t)-(g_1(X)+d_1(t_0))|  \leq |g_1(X)+d_1(t)| + |g_1(X)+d_1(t_0)| \leq 2\psi(X)$
for any $X$. By denoting $\eta_d = 2\psi(0)+d_1(t_0)$, the bound of $d_1(t)$ satisfies that $\sup_{t\geq t_0} |d_1(t)| \leq \eta_d$,
which implies that $\sqrt{\|\bar{\Gamma}_0(t)\|^2 + \|\bar{\Gamma}_1(t)\|^2}$ is bounded:
\begin{equation}\label{proD_20}
\sqrt{\|\bar{\Gamma}_0(t)\|^2 + \|\bar{\Gamma}_1(t)\|^2} \leq \|\bar{\Gamma}_0(t)\|+\|\bar{\Gamma}_1(t)\|
\leq \eta_{\Gamma12},
\end{equation}
for $t\geq t_0$,
where $\eta_{\Gamma12} = (\gamma_{01}+\gamma_{11})(\eta_d+|\bar{D}|)+\eta_d+ \sqrt{n+1} \gamma_{12} \eta_r $.
Therefore,
{\fontsize{9pt}{9pt}\begin{equation}\label{proD_21}
\begin{split}
&\lim_{t\rightarrow \infty} \int_{t_0}^{t} e^{-\bar{\alpha}_1(t-\tau)} \sqrt{\|\bar{\Gamma}_0(\tau)\|^2 + \|\bar{\Gamma}_1(\tau)\|^2} d\tau  \\
\leq& \lim_{t\rightarrow \infty}  \int_{0}^{t/2} e^{-\bar{\alpha}_1(\frac{t}{2}-\tau)}  \sqrt{\| \bar{\Gamma}_0(\tau+\frac{t}{2})\|^2+\|\bar{\Gamma}_1(\tau+\frac{t}{2})\|^2} d\tau \\
& +\lim_{t\rightarrow \infty} e^{-\frac{\bar{\alpha}_1 t}{2}} \eta_{\Gamma12} \int_{t_0}^{t/2} e^{-\bar{\alpha}_1(\frac{t}{2}-\tau)} d\tau \\
=&0.
\end{split}
\end{equation}}

With the combination of (\ref{proD_15})--(\ref{proD_21}), there is
$\lim_{t\rightarrow \infty} \sqrt{\bar{V}(E,\bar{\xi})}
=0$,
which further implies that $\lim_{t\rightarrow \infty} E(t) = \lim_{t\rightarrow \infty} \bar{\xi}(t) = 0$.
Since $\left[\begin{smallmatrix} X(t) - \hat{X}(t) \\ f(X(t),u(t),t) - \hat{f}(t) \end{smallmatrix}\right] = T_1\bar{\xi}(t) + T_1B_2 (d_1(t)-\bar{D})$,
it can be deduced that $\lim\limits_{t\rightarrow \infty} (X(t) - \hat{X}(t)) =  \lim\limits_{t\rightarrow \infty} (f(X(t),u(t),t) - \hat{f}(t))= 0$. Thus, (\ref{Thm_converge_re}) is proved.
\hfill $\square$

%\section*{References}

\bibliographystyle{IEEEtran}
\bibliography{total_xue}

\begin{thebibliography}{10}
\providecommand{\url}[1]{#1}
\csname url@samestyle\endcsname
\providecommand{\newblock}{\relax}
\providecommand{\bibinfo}[2]{#2}
\providecommand{\BIBentrySTDinterwordspacing}{\spaceskip=0pt\relax}
\providecommand{\BIBentryALTinterwordstretchfactor}{4}
\providecommand{\BIBentryALTinterwordspacing}{\spaceskip=\fontdimen2\font plus
\BIBentryALTinterwordstretchfactor\fontdimen3\font minus
  \fontdimen4\font\relax}
\providecommand{\BIBforeignlanguage}[2]{{%
\expandafter\ifx\csname l@#1\endcsname\relax
\typeout{** WARNING: IEEEtran.bst: No hyphenation pattern has been}%
\typeout{** loaded for the language `#1'. Using the pattern for}%
\typeout{** the default language instead.}%
\else
\language=\csname l@#1\endcsname
\fi
#2}}
\providecommand{\BIBdecl}{\relax}
\BIBdecl

\bibitem{kumar2014control}
P.~Kumar, ``Control: A perspective,'' \emph{Automatica}, vol.~50, no.~1, pp.
  3--43, 2014.

\bibitem{Wang2007PIDTAC}
D.~{Wang}, ``Further results on the synthesis of pid controllers,'' \emph{IEEE
  Transactions on Automatic Control}, vol.~52, no.~6, pp. 1127--1132, 2007.

\bibitem{Li2014RobustAdapTAC}
S.~S. {Ge} and Z.~{Li}, ``Robust adaptive control for a class of mimo nonlinear
  systems by state and output feedback,'' \emph{IEEE Transactions on Automatic
  Control}, vol.~59, no.~6, pp. 1624--1629, 2014.

\bibitem{Zhang2015slidingTAC}
X.~{Zhang}, H.~{Su}, and R.~{Lu}, ``Second-order integral sliding mode control
  for uncertain systems with control input time delay based on singular
  perturbation approach,'' \emph{IEEE Transactions on Automatic Control},
  vol.~60, no.~11, pp. 3095--3100, 2015.

\bibitem{Chen2015DOTAC}
M.~{Chen}, P.~{Shi}, and C.~{Lim}, ``Robust constrained control for mimo
  nonlinear systems based on disturbance observer,'' \emph{IEEE Transactions on
  Automatic Control}, vol.~60, no.~12, pp. 3281--3286, 2015.

\bibitem{Chen7265050TIE}
W.~{Chen}, J.~{Yang}, L.~{Guo}, and S.~{Li}, ``Disturbance-observer-based
  control and related methods¡ªan overview,'' \emph{IEEE Transactions on
  Industrial Electronics}, vol.~63, no.~2, pp. 1083--1095, 2016.

\bibitem{freidovich2008performance}
L.~B. Freidovich and H.~K. Khalil, ``Performance recovery of
  feedback-linearization-based designs,'' \emph{IEEE Transactions on Automatic
  Control}, vol.~53, no.~10, pp. 2324--2334, 2008.

\bibitem{khalil2017cascadeAc}
H.~K. Khalil, ``Cascade high-gain observers in output feedback control,''
  \emph{Automatica}, vol.~80, pp. 110--118, 2017.

\bibitem{khalil2017highBOOK}
H.~H. Khalil, \emph{High-gain observers in nonlinear feedback control}.\hskip
  1em plus 0.5em minus 0.4em\relax SIAM, 2017, vol.~31.

\bibitem{han2009pid}
J.~Han, ``From {PID} to active disturbance rejection control,'' \emph{IEEE
  Transactions on Industrial Electronics}, vol.~56, no.~3, pp. 900--906, 2009.

\bibitem{HuangADRCreviewISA}
Y.~Huang and W.~Xue, ``Active disturbance rejection control: Methodology and
  theoretical analysis,'' \emph{ISA Transaction}, vol.~53, pp. 963--976, 2014.

\bibitem{Guo2015ADRCTAC}
B.~{Guo} and H.~{Zhou}, ``The active disturbance rejection control to
  stabilization for multi-dimensional wave equation with boundary control
  matched disturbance,'' \emph{IEEE Transactions on Automatic Control},
  vol.~60, no.~1, pp. 143--157, 2015.

\bibitem{Ohnishi6822593TIE}
E.~{Sariyildiz} and K.~{Ohnishi}, ``Stability and robustness of
  disturbance-observer-based motion control systems,'' \emph{IEEE Transactions
  on Industrial Electronics}, vol.~62, no.~1, pp. 414--422, 2015.

\bibitem{Texas2013}
{Texas Instruments}, \emph{Technical Reference Manual, TMS320F28069M,
  TMS320F28068M InstaSPINTMMOTION Software, Literature Number: SPRUHJ0A.},
  2013.

\bibitem{sun2016tuning}
L.~Sun, D.~Li, K.~Hu, K.~Y. Lee, and F.~Pan, ``On tuning and practical
  implementation of active disturbance rejection controller: a case study from
  a regenerative heater in a 1000 {MW} power plant,'' \emph{Industrial \&
  Engineering Chemistry Research}, vol.~55, no.~23, pp. 6686--6695, 2016.

\bibitem{xue2018performance}
W.~Xue and Y.~Huang, ``Performance analysis of 2-{DOF} tracking control for a
  class of nonlinear uncertain systems with discontinuous disturbances,''
  \emph{International Journal of Robust and Nonlinear Control}, vol.~28, no.~4,
  pp. 1456--1473, 2018.

\bibitem{Zhao7959174Tac}
Z.~{Zhao} and B.~{Guo}, ``A novel extended state observer for output tracking
  of {MIMO} systems with mismatched uncertainty,'' \emph{IEEE Transactions on
  Automatic Control}, vol.~63, no.~1, pp. 211--218, 2018.

\bibitem{guo2015active}
B.~Guo and Z.~Zhao, ``Active disturbance rejection control: Theoretical
  perspectives,'' \emph{Communications in Information and Systems}, vol.~15,
  no.~3, pp. 361--421, 2015.

\bibitem{WANG2016AC}
X.~Wang, S.~Li, and J.~Lam, ``Distributed active anti-disturbance output
  consensus algorithms for higher-order multi-agent systems with mismatched
  disturbances,'' \emph{Automatica}, vol.~74, pp. 30 -- 37, 2016.

\bibitem{Yao7900325TIE}
J.~{Yao} and W.~{Deng}, ``Active disturbance rejection adaptive control of
  hydraulic servo systems,'' \emph{IEEE Transactions on Industrial
  Electronics}, vol.~64, no.~10, pp. 8023--8032, 2017.

\bibitem{Chen2019SCIS}
S.~Chen, W.~Bai, Y.~Hu, Y.~Huang, and G.~Zhiqiang, ``On the conceptualization
  of total disturbance and its profound implications,'' \emph{SCIENCE CHINA
  Information Sciences}, 2019.

\bibitem{GaoESObandwidth}
D.~Yoo, S.~S.-T. Yau, and Z.~Gao, ``Optimal fast tracking observer bandwidth of
  the linear extended state observer,'' \emph{International Journal of
  Control}, vol.~80, no.~1, pp. 102--111, 2007.

\end{thebibliography}

\end{document}